# INFORMATION MEASURES IN DETECTING AND RECOGNIZING SYMMETRIES

*Denis V. Popel*

Department of Computer Science,
Baker University, Baldwin City, KS 66006,
`popel@ieee.org`

**ABSTRACT**

This paper presents a method to detect and recognize symmetries in Boolean functions. The idea is to use *information theoretic measures* of Boolean functions to detect sub-space of possible symmetric variables. Coupled with the new techniques of efficient estimations of information measures on *Binary Decision Diagrams* (BDDs) we obtain promised results in symmetries detection for large-scale functions.

## 1. INTRODUCTION

Determining symmetries among groups of variables is important in logic synthesis [4], design verification and testing [5], as well as in problems of technology mapping, such as Boolean matching [6, 12]. The effectiveness of the matching procedure can be increased if the groups of symmetric variables are known. The symmetry properties are used in different areas of logic design, namely, in decomposition and minimization [3, 8].

There are several techniques to recognize symmetries based on the following principles:

**Manipulation of tabular data.** The well-known algorithms explore the properties of symmetries to manipulate truth tables. For example, in [11], an efficient method to detect different types of symmetries based on numerical methods has been proposed.

**Transformation into spectral domain.** This principle exploits the properties of spectra to determine the symmetries in variables for a given function. The results on detecting symmetries in Hadamard, Haar and other transform bases have been reported [4].

**Formal representation of symmetric functions** (DTs and DDs, Reed-Muller expressions). In recent years, DDs have been used as an efficient data structure in circuit synthesis, and symmetry detection has become feasible for large-scale functions [1, 9].

In our approach, we consider information theoretic measures at the first phase of forming a *sub-space of possible symmetric* variables [8]. At the second phase, we apply the exact (naive) technique to recognize symmetries in the obtained sub-space. DDs are used to calculate efficiently the spectrum of information measures [7].

## 2. THE TYPES OF SYMMETRY

There are four cofactors, namely, $f_{x_i x_j}$, $f_{x_i \overline{x}_j}$, $f_{\overline{x}_i x_j}$ and $f_{\overline{x}_i \overline{x}_j}$, for any pair of variables $\{x_i, x_j\}$ of a Boolean function $f = f(x_1, \ldots, x_i, \ldots, x_j, \ldots, x_n)$ [2].

**Definition 1** *A function $f$ has* non-equivalence symmetry *(NE) in variables $\{x_i, x_j\}$, if it remains invariant when $x_i$ and $x_j$ (or $\overline{x}_i$ and $\overline{x}_j$) are interchanged: $f_{\overline{x}_i x_j} = f_{x_i \overline{x}_j}$.*

**Definition 2** *A function $f$ has* equivalence symmetry *(E) in variables $\{x_i, x_j\}$, if it remains invariant when $x_i$ and $\overline{x}_j$ (or $\overline{x}_i$ and $x_j$) are interchanged: $f_{x_i x_j} = f_{\overline{x}_i \overline{x}_j}$.*

**Definition 3** *A function $f$ has* multiform symmetry *(M) in variables $\{x_i, x_j\}$, if it is simultaneously NE- and E-symmetric in $\{x_i, x_j\}$.*

**Example 1** *The following functions are symmetric:*
*(i) $f = x_2 \oplus x_3 \oplus x_2 x_3 \oplus x_1 x_2 x_3$ is NE-symmetric in $\{x_2, x_3\}$, and $f = x_1 x_3 \vee x_1 x_2$ is NE-symmetric in $\{x_2, x_3\}$; (ii) $f = \overline{x}_1 \oplus x_2 \oplus \overline{x}_1 x_2 x_3$ is E-symmetric in $\{\overline{x}_1, x_2\}$; (iii) $f = x_1 \overline{x}_2 \vee \overline{x}_1 x_2$ is M-symmetric in $\{x_1, x_2\}$.*

A function $f$ is *partially symmetric* with respect to subset $X_t$ of variables, $X_t \subseteq X$, if any permutation of variables in $X_t$ leaves $f$ unchanged. A function $f$ is *totally symmetric* if every pair of variables in the function is either NE- or E-symmetric.

**Example 2** *The following functions are symmetric:*
*(i) $f = \overline{x}_1 \oplus x_2 \oplus \overline{x}_1 x_2 x_3$ is partially symmetric, i.e. $E\{x_1, x_2\}$; (ii) $f = x_1 \overline{x}_2 \vee \overline{x}_1 x_2$ and $f = x_1 x_2 \oplus x_1 x_3 \oplus x_2 x_3$ are totally symmetric.*

We summarize these types of symmetry in Table 1.

Table 1: Types of symmetry

| Name | Condition | Notation |
|---|---|---|
| non-equivalence | $f_{\overline{x}_i x_j} = f_{x_i \overline{x}_j}$ | $\{x_i, x_j\}$ or $NE\{x_i, x_j\}$ |
| equivalence | $f_{x_i x_j} = f_{\overline{x}_i \overline{x}_j}$ | $\{x_i, \overline{x}_j\}$ or $E\{x_i, x_j\}$ |
| multiform | $f_{\overline{x}_i x_j} = f_{x_i \overline{x}_j}$ | $M\{x_i, x_j\}$ |
| | $f_{x_i x_j} = f_{\overline{x}_i \overline{x}_j}$ | |

## 2.1. Information theory notations

In order to quantify the content of information for a finite field of events $A = \{a_1, a_2, \cdots, a_n\}$ with probabilities distribution $\{p(a_i)\}$, $i = 1, 2, \cdots, n$, Shannon introduced the concept of *entropy* [10]: $H(A) = -\sum_{i=1}^{n} p(a_i) \cdot \log p(a_i)$, where log denotes the base 2 logarithm. For two finite fields of events $A$ and $B$ with probability distribution $\{p(a_i)\}, i = 1, 2, \cdots, n$, and $\{p(b_j)\}$, $j = 1, 2, \cdots, m$, probability of the joint occurrence of $a_i$ and $b_j$ is joint probability $p(a_i, b_j)$, and there is conditional probability, $p(a_i|b_j) = p(a_i, b_j)/p(b_j)$. The *conditional entropy* of $A$ given $B$ is defined by $H(A|B) = -\sum_{i=1}^{n} \sum_{j=1}^{m} p(a_i, b_j) \cdot \log p(a_i|b_j)$.

**Example 3** *Let us calculate the entropy of a Boolean function $f$ given by its truth column vector* [1100000111000010]: $H(f) = -^6/_{16} \log_2 {}^6/_{16} - {}^{10}/_{16} \log_2 {}^{10}/_{16} = 0.95$ *bit/pattern. The conditional entropy with respect to variable $x_1$ is* $H(f|x_1) = -{}^5/_{16}(\log_2 {}^5/_{16} - \log_2 {}^8/_{16}) - {}^3/_{16}(\log_2 {}^3/_{16} - \log_2 {}^8/_{16}) - {}^5/_{16}(\log_2 {}^5/_{16} - \log_2 {}^8/_{16}) - {}^3/_{16}(\log_2 {}^3/_{16} - \log_2 {}^8/_{16}) = 0.95$ *bit/pattern.*

## 3. INFORMATION THEORETIC MEASURES IN SYMMETRY DETECTION

In this Section, we focus on detecting different types of symmetries ($NE$-, $E$-, $M$- and total symmetries) by information theoretic measures via design of decision diagram.

### 3.1. Detection of Non-equivalent Symmetry

**Lemma 1** *A Boolean function $f$ is $NE$-symmetric in $\{x_i, x_j\}$, if $f_{\overline{x}_i} = f_{\overline{x}_j}$ and $f_{x_i} = f_{x_j}$.*

**Theorem 2** *If a Boolean function $f$ is $NE$-symmetric in $\{x_i, x_j\}$, then $H(f|x_i) = H(f|x_j)$ and $H(f_{x_i}) = H(f_{x_j})$, $H(f_{\overline{x}_i}) = H(f_{\overline{x}_j})$.*

**Theorem 3** *The condition $H(f|x_i) = H(f|x_j)$ from the Theorem 2 is necessary but not sufficient to detect $NE$-symmetry.*

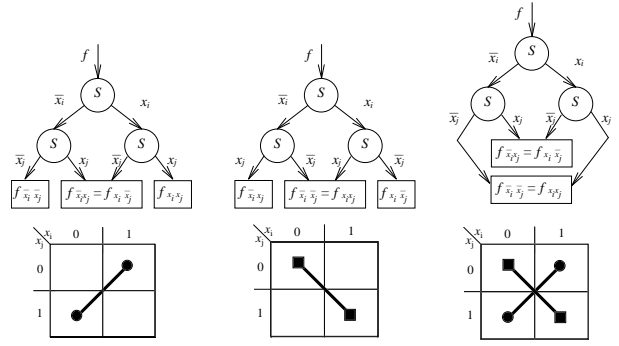

Figure 1: Parts of BDD (nodes primitives) of a function $f$ that is $NE$-, $E$- and $M$-symmetric in variables $x_i, x_j$

**Example 4** *Consider a Boolean function $f$ of four variables specified by the truth column vector* [1100000111000010]. *The entropy measures and cofactors of the function are given in tables below.*

| | $H(f_{\overline{x}})$ | $H(f_x)$ | | $f_{\overline{x}}$ | $f_x$ |
|---|---|---|---|---|---|
| $x_1$ | 0.95 | 0.95 | $x_1$ | [11000010] | [11000001] |
| $x_2$ | **0.81** | *1* | $x_2$ | [00010010] | [11001100] |
| $x_3$ | **0.81** | *1* | $x_3$ | [00010010] | [11001100] |
| $x_4$ | 0.95 | 0.95 | $x_4$ | [10011000] | [10001001] |

*We analyze the pairs of variables for which the information measures take equal values. The analysis shows that $NE$-symmetries in $\{x_2, x_3\}$ and in $\{x_1, x_4\}$ are possible.*

*The function $f$ is $NE$-symmetric in $\{x_2, x_3\}$. It means that $f_{\overline{x}_2} = f_{\overline{x}_3}$. This function is $NE$-symmetric in $\{x_1, x_4\}$ as well. Taking into consideration the permutation of variables assignments, we obtain $f_{\overline{x}_1} = f_{\overline{x}_4}$. As a result of BDD design, we obtain the following expression: $f = \overline{x}_2 \cdot \overline{x}_3 \vee \overline{x}_1 \cdot x_2 \cdot x_3 \cdot x_4 \vee x_1 \cdot x_2 \cdot x_3 \cdot \overline{x}_4$.*

### 3.2. Detection of Equivalent Symmetry

It is easy to show that a Boolean function $f$ is $E$-symmetric in $\{x_i, \overline{x}_j\}$, if $f_{\overline{x}_i} = f_{x_j}$ and $f_{x_i} = f_{\overline{x}_j}$. $E$-symmetry condition $f_{x_i x_j} = f_{\overline{x}_i \overline{x}_j}$ implies $f_{x_i} = f_{\overline{x}_j}$ and $f_{\overline{x}_i} = f_{x_j}$. Nodes with $E$-symmetric variables are placed together through design of DDs (Figure 1).

**Property 1** *If a Boolean function $f$ is $E$-symmetric in $\{x_i, \overline{x}_j\}$, then $H(f|x_i) = H(f|x_j)$ and $H(f_{x_i}) = H(f_{\overline{x}_j})$, $H(f_{\overline{x}_i}) = H(f_{x_j})$.*

**Remark 1** *Property 1 is necessary, but not sufficient to detect $E$-symmetry.*

**Example 5** *Consider a Boolean function $f$ of three variables given by its truth column vector $[11100011]$. The information measures and cofactors for this function are shown below.*

|       | $H(f_{\overline{x}})$ | $H(f_x)$ |       | $f_{\overline{x}}$ | $f_x$  |
|-------|----------|----------|-------|--------|--------|
| $x_1$ | **0.81** | 1        | $x_1$ | [1110] | [0011] |
| $x_2$ | 1        | 0.81     | $x_2$ | [1100] | [1011] |
| $x_3$ | 0.81     | 1        | $x_3$ | [1101] | [1001] |

*Following information theoretic measures, we expect E-symmetry in $\{x_1, \overline{x}_2\}$ and in $\{x_2, \overline{x}_3\}$. Really, the function $f$ is E-symmetric in $\{x_1, \overline{x}_2\}$, because $f_{\overline{x}_1} = f_{x_2}$ and $f_{x_1} = f_{\overline{x}_2}$. But it is not E-symmetric in $\{x_2, \overline{x}_3\}$ because $f_{\overline{x}_2} \neq f_{x_3}$. The function can be represented by the following AND/OR expression: $f = \overline{x}_1 \cdot \overline{x}_2 \vee x_1 \cdot x_2 \vee \overline{x}_1 \cdot x_2 \cdot \overline{x}_3$.*

### 3.3. Detection of Multiform and Totally Symmetries

**Property 2** *If a Boolean function $f$ is $M$-symmetric in $\{x_i, x_j\}$, then $H(f|x_i) = H(f|x_j)$ and $H(f_{x_i}) = H(f_{x_j}) = H(f_{\overline{x}_i}) = H(f_{\overline{x}_j})$.*

A part of the DT (nodes primitive) to be constructed for a function $f$ that is $M$-symmetric in variables $x_i, x_j$ is given in Figure 1.

**Example 6** *(Continuation of Example 4) The function $f$ is $M$-symmetric in $\{x_1, x_4\}$, because $H(f_{x_1}) = H(f_{x_4}) = H(f_{\overline{x}_1}) = H(f_{\overline{x}_4}) = 0.95$ bit/pattern.*

**Property 3** *If a Boolean function $f$ is totally symmetric, then $H(f_{x_1}) = H(f_{\overline{x}_1}) = \ldots = H(f_{x_n}) = H(f_{\overline{x}_n})$.*

**Example 7** *Consider a Boolean function $f$ of three variables given by its truth column vector $[00010111]$. The entropy measures and cofactors for the function are presented in the tables below.*

|       | $H(f_{\overline{x}})$ | $H(f_x)$ |       | $f_{\overline{x}}$ | $f_x$  |
|-------|----------|----------|-------|--------|--------|
| $x_1$ | **0.81** | 0.81     | $x_1$ | [0001] | [0111] |
| $x_2$ | 0.81     | 0.81     | $x_2$ | [0001] | [0111] |
| $x_3$ | 0.81     | 0.81     | $x_3$ | [0001] | [0111] |

*The information theoretic measures are equal, so the function is totally symmetric: $f = \overline{x}_1 \cdot x_2 \cdot x_3 \vee x_1 \cdot \overline{x}_2 \cdot x_3 \vee x_1 \cdot x_2 \cdot \overline{x}_3 \vee x_1 \cdot x_2 \cdot x_3$.*

### 3.4. BDD Based Technique for Calculation of Information Measures

Our technique for exact computation of information measures for Boolean functions represented in the form of BDDs exploits the following. The conditional entropy $H(f|x)$ of the function $f$ with respect to the variable $x$ can be simplified using the theorem below:

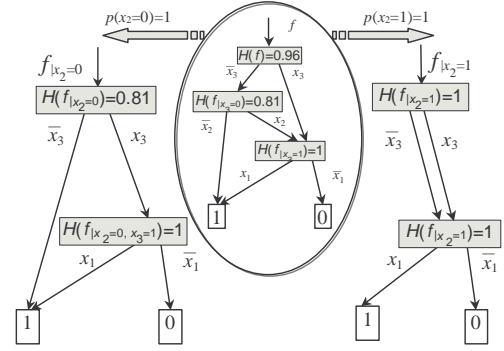

Figure 2: Computing information measures using BDD

**Theorem 4** *The conditional entropy $H(f|x)$ can be calculated by the following equation:*

$$H(f|x) = p(x=0) \cdot H(f_{|x=0}) + p(x=1) \cdot H(f_{|x=1})$$

It means that for calculation of conditional entropy we need to compute the entropy of each sub-function. In this case probability must be assigned to every node in BDD in order to distribute the desired output probability to the root.

**Example 8** *The entropy of the function $f = \overline{x}_3 \cdot \overline{x}_2 \vee x_1$ be: $H(f) = 0.96$ bit. The conditional entropy of the function $f$ given $x_2$ be: $H(f|x_2) = {}^1\!/_2 \cdot H(f_{|x_2=0}) + {}^1\!/_2 \cdot H(f_{|x_2=1}) = 0.41 + 0.5 = 0.91$ bit (Figure 2). The same manipulation yields: $H(f|x_1) = 0.41$ bit and $H(f|x_3) = 0.91$ bit. The conditional entropy of the function $f$ given a set of variables $\{x_1, x_2\}$ be: $H(f|x_1 x_2) = 0.25$ bit.*

## 4. ALGORITHM AND EXPERIMENTAL RESULTS

We propose an algorithm to detect symmetries of Boolean functions using BDDs, called $InfoRECSym - DD$ ($Inf o$rmation $REC$ognizer of $Sym$metries). A sketch of the algorithm is given in Figure 3.

The original program presented in [8] was modified to detect possible symmetries groups using binary decision diagrams representation. Table 2 contains a fragment of our results comparing to the strategy published by Tsai et al. [11]. We use the notation $(S,N)$, where $S$ means the size of a symmetric group (the number of symmetric variables), and $N$ means the number of symmetric groups.

## 5. CONCLUDING REMARKS

This paper addresses a method for detecting and recognizing different types of symmetries (totally, partially $NE$, $E$, $M$) in Boolean functions. The method is based on a variety of information measures computed on decision diagrams.

Table 2: Results of $InfoRecSym - DD$ in symmetry detection

|  | | Tsai et al. [11] | | $InfoRecSym - DD$ | |
|---|---|---|---|---|---|
|  | I/O nr. | $(S, N)$ | $Time$ | $(S, N)$ | $Time$ |
| cm82 | 5/1 | (3,1) (2,2) | 0.044 | (3,1) (2,2) | 0.00 |
| f51m | 5/3 | (2,2) | 0.113 | (4,1) | 0.113 |
| z4ml | 7/1 | (3,1) (2,2) | 0.113 | (3,1) (2,2) | 0.00 |
| x4 | 10/66 | (4,1) (3,1) (2,1) | 0.275 | (8,1) (4,1) | 0.13 |
| x3 | 13/13 | (10,1) (2,1) | 0.715 | (10,1) (2,1) | 1.05 |
| apex6 | 13/41 | (10,1) (2,1) | 0.691 | (10,1) (2,1) | 0.48 |
| des | 18/120 | (4,1) (2,7) | 3.602 | (13,1) (3,1) (17,1) | 11.16 |
| apex7 | 19/8 | (8,1) (3,1) (2,2) | 4.917 | (8,1) (3,1) (2,2) | 0.02 |

```
Input Decision Diagram of a function f
Output The pairs of symmetric variables {x_i,x_j}
with detected types of symmetries
```

$InfoRECSym - DD(f)$
{
  **for**(All pairs $\{x_i, x_j\}$)
    **if** $(H(f|x_i) = H(f|x_j))$
    {
      **if**$(H(f_{\overline{x}_i}) = H(f_{x_j}))$
        **if** $(f_{\overline{x}_i} = f_{\overline{x}_j}$ and $f_{x_i} = f_{x_j})$
          **then** $f$ is $NE$-symmetric in $\{x_i, x_j\}$
      **if**$(H(f_{\overline{x}_i}) = H(f_{x_j}))$
        **if** $(f_{\overline{x}_i} = f_{x_j}$ and $f_{x_i} = f_{\overline{x}_j})$
          **then** $f$ is $E$-symmetric in $\{x_i, \overline{x}_j\}$
      **if** ($f$ is $NE$-, $E$-symmetric simultaneously)
        **then** $f$ is $M$-symmetric in $\{x_i, x_j\}$
    }
  **if**(All pairs are either $NE$- or $E$-symmetric)
    **then** $f$ is totally symmetric function.
}

Figure 3: Sketch of the algorithm $InfoRECSym - DD$

## 6. REFERENCES

[1] J. Butler and T. Sasao. On the properties of multiple-valued functions that are symmetric in both variables and labels. In *Proc. IEEE Int. Symp. on Multiple-Valued Logic*, pages 236–241, 1998.

[2] S. Das and C. Sheng. On detecting total or partial symmetry of switching functions. *IEEE Trans. on Computers*, pages 352–355, 1971.

[3] R. Drechsler and B. Becker. *Binary Decision Diagrams: theory and implementation*. Kluwer Academic Publishers, 1999.

[4] C. Edwards and S. Hurst. A digital synthesis procedure under function symmetries and mapping methods. *IEEE Trans. on Computer*, C-27(11):985–997, 1978.

[5] W. Ke and P. Menon. Delay-testable implementation of symmetric functions. *IEEE Trans. on Computers*, C-14(6):772–775, 1995.

[6] I. Pomeranz and S. Reddy. On determining symmetries in inputs of logic circuits. *IEEE Trans. on CAD of Integrated Circuits and Systems*, 13(11):1478–1433, 1994.

[7] D. Popel. Towards efficient calculation of information measures for reordering of binary decision diagrams. In *Proc. IEEE Int. Symp. on Signals, Circuits and Systems*, pages 509–512, 2001.

[8] D. Popel, P. Dziurzanski, and A. Tomaszewska. On detecting symmetries in switching functions using entropy measures. In *Proc. Int. Conf. on Advanced Computer Systems - ACS'99*, pages 475–482, 1999.

[9] C. Scholl, D. Moller, P. Molitor, and R. Drechsler. BDD minimization using symmetries. *In IEEE Trans. on CAD of Integrated Circuits and Systems*, 18(2):81–100, 1999.

[10] C. Shannon. A mathematical theory of communication. *Bell Syst. Tech. J.*, 27:379–423, 623–656, 1948.

[11] C. Tsai and M. Marek-Sadowska. Generalized Reed-Muller forms as a tool to detect symmetries. *IEEE Trans. on Computers*, C-45(1):33–40, 1996.

[12] K. Wang and T. Hwang. Boolean matching for incompletely specified functions. *IEEE Trans. on CAD of Integrated Circuits and Systems*, 16(2):160–168, 1997.